# Tuning of impurity-bound interlayer complexes in a van der Waals heterobilayer

Fabien Vialla[1,2], Mark Danovich[3], David A. Ruiz-Tijerina[3], Mathieu Massicotte[1], Peter Schmidt[1], Takashi Taniguchi[4], Kenji Watanabe[4], Ryan J. Hunt[5], Marcin Szyniszewski[5], Neil D. Drummond[5], Thomas G. Pedersen[6,7], Vladimir I. Fal'ko[3,8], Frank H.L. Koppens[1,9]

[1]ICFO–Institut de Ciències Fotòniques, The Barcelona Institute of Science and Technology, Castelldefels, Barcelona 08860, Spain. [2]Institut Lumière Matière UMR 5306, Université Claude Bernard Lyon 1, CNRS, Université de Lyon, F-69622 Villeurbanne, France. [3]National Graphene Institute, University of Manchester, Booth St E, Manchester M13 9PL, United Kingdom. [4]National Institute for Materials Science, 1-1 Namiki, Tsukuba 305-0044, Japan. [5]Department of Physics, Lancaster University, Lancaster LA1 4YB, United Kingdom. [6]Department of Physics and Nanotechnology, Aalborg University, DK-9220 Aalborg East, Denmark. [7]Center for Nanostructured Graphene (CNG), DK-9220 Aalborg Øst, Denmark. [8]Henry Royce Institute for Advanced Materials, M13 9PL, Manchester, United Kingdom. [9]ICREA – Institució Catalana de Recerça i Estudis Avancats, 08010 Barcelona, Spain.

**Due to their unique two-dimensional nature, charge carriers in semiconducting transition metal dichalcogenides (TMDs) exhibit strong unscreened Coulomb interactions and sensitivity to defects and impurities. The versatility of van der Waals layer stacking allows spatially separating electrons and holes between different TMD layers with staggered band structure, yielding interlayer few-body excitonic complexes whose nature is still debated. Here we combine quantum Monte Carlo calculations with spectrally and temporally resolved photoluminescence measurements on a top- and bottom-gated $MoSe_2/WSe_2$ heterostructure, and identify the emitters as impurity-bound interlayer excitonic complexes. Using independent electrostatic control of doping and out-of-plane electric field, we demonstrate control of the relative populations of neutral and charged complexes, their emission energies on a scale larger than their linewidth, and an increase of their lifetime into the microsecond regime. This work unveils new physics of confined carriers and is key to the development of novel optoelectronics applications.**

Transition metal dichalcogenides form a new class of semiconducting two-dimensional (2D) materials that display strongly bound electron-hole complexes [1-4] — such as excitons and trions — and original valley physics [5-8] up to room temperature. Furthermore, their unique layered structure allows them to be stacked in van der Waals heterostructures with atomically sharp and clean interfaces [9-11]. The combination of those novel properties with the versatility of layer engineering opens up new avenues for room-temperature excitonic complex formation and manipulation [12-14]. In particular, the model case of TMD heterobilayers with type-II (staggered) band alignment exhibits, upon light excitation, ultrafast charge transfer between layers [15-17] and luminescence at energies lower than the one generated by intralayer complexes [18-22]. This is interpreted as the formation of interlayer excitons where electrons and holes reside in different layers due to the staggered band alignment [22-26]. Promising properties of those emitters have been demonstrated, such as long lifetime [20,27-29] and long spin-valley population time [29,30], establishing them as good candidates to achieve coherent manipulation [31-34]. Yet, clear identification of the nature of these emitters and *in situ* control over them are crucial challenges that still need to be overcome.

In the case of individual monolayer TMDs, electrostatic tuning of the carrier density using a single gate has been key to unveiling the competition between delocalized excitonic complexes (neutral and charged excitons), as well as the presence of impurity-bound local states [5,35,36]. However, in heterobilayer junctions this standard approach is insufficient to achieve unambiguous identification and control over the complexes [20,37,38]. This can be understood considering the spatial separation between electrons and holes residing in different layers. With carriers further apart, interlayer complexes present much longer recombination lifetimes, as compared to their intralayer counterparts [27,28], hence interlayer delocalized carriers and/or complexes are more likely to diffuse, interact and bind with charged impurities, thus forming localized complexes. Furthermore, stable interlayer complexes should present a permanent out-of-plane dipole, introducing a strong dependence on the local electric field. Therefore, thorough characterization and control over the complete local electrostatic environment, charge and field distributions, are required to reliably investigate excitonic complexes in this type of system.

Here, we make use of a dual-gated heterostructure design that allows independent control over the charge carrier density and out-of-plane electric field in a TMD heterobilayer junction. Tuning the carrier density modifies the luminescence intensity and lineshape, corresponding to a change in the relative populations of the stable complexes in the junction. Remarkably, we observe that the charged complexes present energies larger than the neutral ones, in contrast to the monolayer case. With the support of *ab initio* calculations [39], we unveil the localized, impurity-bound nature of the interlayer complexes yielding the strongest luminescence lines. This observation emphasizes the crucial role of defects in such confined system. Furthermore, varying the electric field, we demonstrate control over a large range of the properties of these complexes such as their emission energy (due to the linear Stark effect) and their lifetime, enabling *in situ* manipulation of the complexes.

We fabricated a van der Waals heterostructure based on stacked $MoSe_2/WSe_2$ monolayers encapsulated in multilayer hexagonal boron nitride (hBN), using a high-temperature flake transfer technique [40] (see Figure 1a for a schematics and Figure 1b for an optical micrograph). The two TMD monolayers are aligned following their straight edges and contacted with evaporated Ti/Au leads before complete hBN encapsulation. The substrate, made of highly doped Si below a 285 nm-thick $SiO_2$ layer, acts as a global bottom gate at potential $V_b$, while a thin 10 nm gold film deposited on top of the hBN is used as a transparent local top gate at potential $V_t$. The static electric field in the device is therefore entirely defined by the voltages applied to both gates. We quantitatively evaluate the out-of-plane field $F(V_t,V_b)$ at the individual monolayers and their junction, from analytical and finite-element models (see SI S1). Effects of the potentials on the electrostatically induced change in carrier density $n(V_t,V_b)$ of the flakes can also be extracted (see SI S1), allowing us to experimentally disentangle the response of the system to both the electric field and carrier density.

We performed spatially resolved photoluminescence (PL) spectroscopy at low temperature (30 K) to probe the formation of excitonic complexes in the two TMDs and their junction (Figure 1c). First, we characterized each of the isolated monolayer flakes individually, away from the junction (purple and green solid lines in Figure 1c), by monitoring the changes in their respective PL spectra induced by varying the carrier density (see SI S2). We observe the typical behaviour of delocalized complexes in TMD monolayers [35], and identify the intralayer exciton (at 1.64 eV for $MoSe_2$ and 1.69 eV for $WSe_2$) and trion (at 1.61 eV and 1.66 eV, respectively) lines, as well as a low energy tail related to complexes bound to localized impurities [35,36] (around 1.55 eV and 1.6 eV, respectively; see the logarithmic scale plot in the bottom panel of Figure 1c, and the schematics of the corresponding complexes in Figure 1d). We determine the carrier density yielding the strongest emission from the neutral exciton and impurity-bound peaks for each flake (Figure 1e). It corresponds to the configuration where the electrostatically injected carriers compensate the intrinsic doping [35]. We thereby evaluate the intrinsic impurity density of each flake ($n_{MoSe2}$ and $n_{WSe2}$, with the convention $n_{XSe2} > 0$ for donor type impurities). We find in our device that $MoSe_2$ and $WSe_2$ monolayers are respectively n- and slightly p-doped ($n_{MoSe2} \sim 5.10^{12}$ cm$^{-2}$ and $n_{WSe2} \sim -2.10^{11}$ cm$^{-2}$). We note that these carrier densities, for which neutrality is obtained, are virtually independent of the applied out-of-plane field $F$ (see SI S2). This is expected from the planar nature of the intralayer quasiparticles and further validates our disentangled study of $n$ and $F$.

We now focus our attention on the interlayer complexes by studying the PL emitted at the $MoSe_2/WSe_2$ junction, spatially resolved in Figure 1b and spectrally resolved in Figure 1c (blue line). We observe a quenching of both of the intralayer lines and the emergence of a spatially uniform low energy peak (between 1.25 and 1.4 eV), as previously reported in the literature [20,22,26]. We can correctly reproduce the peak lineshape, over the full range of applied potentials ($V_t,V_b$), by introducing an empirical 3-Gaussian-line fit (dashed lines in Figure 1c). In analogy with the previous identification of intralayer complexes in monolayers, the multi-component spectrum reflects the fact that several stable interlayer complexes are formed at the junction upon light excitation. Due to the type-II band alignment between $MoSe_2$ and $WSe_2$ [24,26], these complexes comprise one or several electrons in the $MoSe_2$ K-point conduction-band edge and one or several holes in the $WSe_2$ K-point valence-band edge. During their long lifetime reported e.g. in [20,27-29], photogenerated carriers can diffuse and interact with charged impurities, and relax into their most stable excitonic configuration before radiative recombination. Our approach is therefore to theoretically consider several different configurations for interlayer complexes, both delocalized and (localized) impurity-bound, and identify the most stable ones.

We developed an *ab initio* approach, where the wavefunctions of different complexes are obtained using diffusion (DMC) and variational (VMC) quantum Monte Carlo calculations. As the precise nature of the donor impurities is unclear, we consider only the long-range part of the electrostatic interaction and take into account interlayer Coulomb interactions, polarization effects and screening from the hBN encapsulation, by considering a bilayer generalisation of the Keldysh potential [39,41]. Following the experimental characterization of the intrinsic doping of both flakes ($n_{MoSe2} > 0$ and $|n_{WSe2}| \ll |n_{MoSe2}|$, Figure 1e), we introduced impurities in the model assuming that only positively charged donors (noted $D^+$) are present in the sample, and only in the MoSe$_2$ layer. We evaluated the binding energies and radiative decay rates for the simplest few-carrier excitonic complexes, both delocalized and donor-bound, by numerically solving the corresponding few-body problem in the limit of small relative angle between MoSe$_2$ and WSe$_2$ lattices, which is experimentally provided by the alignment of the flake edges (see SI S3).

From this theoretical approach, we identified the two most stable complexes in the system, noted $D^0h$ and $D^0X$, shown in Figure 1d in real space and in Figure 1f in band structure schematics. We emphasize here the role of the intrinsic impurities in the system since both of these complexes comprise donor-bound electrons in the MoSe$_2$ layer (bound $D^+ + e$, noted $D^0$). In the first case ($D^0h$), these neutralized donors $D^0$ can recombine with free holes $h$ in the WSe$_2$ layer, yielding photon emission at energy $E_{D^0h} = E_g - \mathcal{E}^b_{D^0h}$ = 1.337 eV (central panel of Figure 2a), where $E_g$ = 1.500 eV is the interlayer bandgap between MoSe$_2$ conduction and WSe$_2$ valence bands (assumed here to fit the data, in good agreement with reported values [20,24,26]) and $\mathcal{E}^b_{D^0h}$ = 163 meV is the calculated binding energy of the complex. We note that the energy to dissociate $h$ from $D^0$ is small ($\mathcal{E}^b_{D^0h} = \mathcal{E}^b_{D^0}$ + 0.2 meV where $\mathcal{E}^b_{D^0}$ is the calculated binding energy of the neutral donor only), leaving the holes free even at the considered cryogenic temperature. In the latter case, we refer to the complexes $D^0X$ as donor-bound trions to emphasize the additional electron compared to $D^0h$. However, our notation reflects their main dissociation channel, which is into a neutral donor $D^0$ and a delocalized interlayer exciton $X$. The complex $D^0X$ can radiatively recombine leaving behind a neutral donor $D^0$, and therefore emitting a photon at energy $E_{D^0X} = E_g + \mathcal{E}^b_{D^0} - \mathcal{E}^b_{D^0X}$ = 1.385 eV (right panel of Figure 2a), where $\mathcal{E}^b_{D^0X}$ = 278 meV is the calculated $D^0X$ total binding energy. Note that, although the absolute energy of the $D^0X$ complex is lower than that of the neutral donor, the $D^0X$ line appears at higher energy (by approximately 48 meV) compared to the $D^0h$ line. This is caused by the specific decay process into a strongly bound $D^0$ state, whose binding energy promotes a larger emitted photon energy. Additionally, we calculated the radiative recombination of each complex mediated by LO phonon emission, noted as

$D^0h + ph$ (left panel of Figure 2a) and $D^0X + ph$. This process yields an additional luminescence line 33 meV below each corresponding main line. We estimated a $D^0h + ph$ line of comparable intensity to the main luminescence peaks, and a much weaker $D^0X + ph$ signal which overlaps with the main $D^0h$ line and can be ignored (see SI S4).

To assign the processes to the respective observed PL lines, we experimentally varied both $V_t$ and $V_b$, such that we measure the emission spectrum for a range of carrier densities, while keeping the out-of-plane field constant (see Figure 2c). We observe that the carrier density mostly affects the relative amplitude between the lines, with minor redshift at large doping that can be attributed to screening of Coulomb interactions by free carriers. Strikingly, we find that the higher energy peak increases in intensity for higher n-type doping. Considering only delocalized complexes does not allow to reproduce the observed doping dependence: In the same way as in the case of delocalized intralayer complexes, calculations show that delocalized interlayer charged trions present a lower energy than delocalized neutral excitons (see SI S5). In contrast, the modelled localized charged complex $D^0X$ appears at higher energy than neutral $D^0h$, consistent with our experimental observations.

We corroborate this peak assignment by comparing our data to the calculated dependence of the emission intensities on the carrier density, as shown in Figures 2b and 2c. We model the relative populations of complexes assuming low thermally and photogenerated electron and hole densities (experimentally <5x10$^{11}$ cm$^{-2}$), compared to the electrostatically injected carrier densities (experimentally >10$^{12}$ cm$^{-2}$). Considering our experimental non-resonant excitation scheme, we estimate that unbound WSe$_2$ holes and MoSe$_2$ electrons are photogenerated in the system, which are able to diffuse in their respective layer and, for the latter, can bind to impurities prior to any recombination. Thus, when electrons are electrostatically removed from the system (-$n_{MoSe2}$ < $n$ < 0), all photogenerated electrons get bound to the available donor impurities to form neutral donors $D^0$. This leads to the preponderance of $D^0h$ complexes in the system, and hence stronger emission from the two low-energy lines attributed to their phonon-mediated and direct recombination, $D^0h + ph$ and $D^0h$. In contrast, when all donors are neutralized by electrostatic electrons (0 < $n$ < $n_{MoSe2}$), photogenerated electrons and holes will bind to the already formed $D^0$, and start forming $D^0X$ complexes at the expense of the $D^0h$. To produce a quantitative analysis of the emission intensities, we model non-radiative recombination as both impurity-driven and Auger processes by introducing the non-radiative cross section $\sigma_{NR}$ for holes to recombine with electrons bound at donor sites (see insets of Figure 2d and SI S6). By combining in steady-state conditions the corresponding non-radiative rates with the radiative recombination rates calculated from the MC approach, we obtain theoretical curves for the doping dependence of the relative peak intensities (lines in Figure 2d) that follow well our experimental data (dots). Deviations from the model can be attributed to the finite temperature and photodoping. In regards to photodoping, we find that increasing excitation power promotes the $D^0X$ line against the one attributed to $D^0h$ (see SI S7). We relate this observation to the high injection of free electrons in MoSe$_2$ with light, consistent with our model.

Next, we fix the carrier density and observe the luminescence of the different peaks while varying the out-of-plane electric field $F$. The independent tuning of $F$ leads to a clear change of the emission energies (Figure 3a). We observe only minor variations in the lineshape over the experimental $F$-range, indicating that the population ratio between complexes is not significantly modified by $F$. In strong contrast, we demonstrate a large linear tuning of the emission energy, over a range of 100 meV, which is larger than the linewidth. We note that all three peaks attributed to the different complex recombinations show the same linear behaviour over the entire carrier density range investigated (see Figure 3b and SI S9). We interpret this effect as a linear Stark shift resulting from the permanent dipole $\mu = e.d = 0.65$ e.nm of the interlayer complexes, where $e$ is the elementary charge and $d$ the interlayer distance. This can be taken into account in our model by considering a field-dependent interlayer bandgap $E_g(F) = E_g(0) + \mu F$. We experimentally extract a value $\mu = 0.9 \pm 0.1$ e.nm where the small discrepancy could be explained by the difficulty of estimating the TMDs effective dielectric constant. This result confirms the spatial structure of the interlayer complexes, where electrons and holes reside in different layers. The observed tuning range is orders of magnitude larger than the ones reported for TMD intralayer exciton lines [42-44], and similar, yet slightly larger, to the one recently observed for interlayer excitons in homobilayer structures [45]. We finally note that a similar design with only thin hBN layers as top and bottom dielectric spacers would yield a shift as high as 500 meV before reaching the breakdown of hBN [46]. This large electrostatic tuning shows great promises for optical modulators based on interlayer complexes.

We further characterized the interlayer emitters by measuring their integrated luminescence decay over time. We observe, as expected [20,27-29], a slow decay on timescales larger than tens of nanoseconds, and originally unveil the strong influence on the effective decay time of both carrier density (Figure 4a) and out-of-plane electric field (Figure 4b). While the decay at long timescales shows an exponential behaviour, and can be related to a lifetime $\tau$, we find a significantly faster decay at short timescales that is more pronounced for larger photon fluxes (see SI S10). In analogy with studies on mono- and multilayer TMDs [47,48], we attribute this behaviour to fast many-body processes that remove electron-hole pairs before complex formation, and effectively introduce a non-linear term in the decay expression (see SI S9). By fitting an analytical solution for the decay over time (lines in Figures 4a,b) to the experimental data, we extract values of $\tau$ that are consistent for all the light fluencies investigated (points in Figure 4c).

We demonstrate that tuning both $n$ and $F$ can strongly modify $\tau$ over orders of magnitude, up to values larger than 0.5 μs which corresponds to our experimental resolution. The strong $F$ dependence can stem from a polarization of the carrier orbitals in the different layers, pulled closer at positive fields and further apart at negative fields, yielding shorter and longer lifetimes respectively (see SI S10). We also find a significant reduction of $\tau$ at large $n$ that can be explained by the increase in non-radiative recombination events, in good agreement with the previously discussed decay rate model (dashed line in figure 4c). The corresponding non-radiative hole cross section $\sigma_{NR}$ is extracted for different fields, and is reported in Figure 4d. Looking back at the steady-state, the emission intensity is governed by the quantum yield, defined as the ratio between radiative and total recombination rates (or likewise the corresponding hole cross sections, see formula in the lower panel of Figure 4d and SI S10). Fixing the $\sigma_{NR}(F)$ values in the fitting of the theoretical quantum yield to the experimental PL intensities, in the low carrier density regime (-$n_{MoSe2}$ < n < 0), we can extract the radiative cross section (sum of $\sigma_{D^0h}(F)$ and $\sigma_{D^0h+ph}(F)$, relative to the $D^0h$ and $D^0h + ph$ emissions respectively, see Figure 4d). Here, we emphasize that we obtain an understanding that is fully consistent with both spectrally and temporally resolved measurements. In particular, our study indicates that reaching neutrality in MoSe$_2$ (n ~ -$n_{MoSe2}$) with large negative field yields interesting experimental conditions with a lifetime in the microsecond regime and the largest quantum yield. Our results can be used as the groundwork for further theoretical studies to achieve quantitative description of the mechanisms governing the field dependence in these original complexes.

In conclusion, we demonstrated, using electrostatic control of interlayer complexes in TMD, a large tuning of many of their properties that opens opportunities for studies as diverse as strong-coupling with cavities, trapping, lasing and condensation up to room temperature [49-51]. We also revealed the dominant role that intrinsic impurities play in the interlayer luminescence spectrum. This shows that crystal purity has to be considered carefully when developing novel optoelectronics applications that would rely on the large scale diffusion or electrostatic funnelling of delocalized interlayer complexes [34]. Finally, we note that the impurity-bound nature of the emitting interlayer complexes should lead to single photon emission at low impurity density, with possible entanglement of the polarisation of the emitted photon with the spin-valley state of the electron left on the donor. In general, our clear understanding and unprecedented electrical control of the excitonic complexes generated in TMD heterobilayers, together with recent advances in valleytronics, introduces promising guidelines for future innovation in quantum optoelectronics.

## METHODS:

**Sample preparation.**
Transition metal dichalcogenide monolayers are obtained from mechanical exfoliation of bulk materials (*HQ Graphene*) using polydimethylsiloxane (PDMS) sheets (*Gel-Pak*). Dry transfer transfer of the isolated flakes is performed using a PDMS/Polypropylene carbonate (*SigmaAldrich*) transparent stamp deposited on a glass slide. Contact between a flake and the stamp in order to pick up the former is done at a temperature of 90°C. Release of a flake from the stamp onto a substrate is done at a temperature of 110°C. Metallic contacts are patterned using laser lithography on a photoresist prior to chemical development and metal evaporation.

**Experimental set-up and data acquisition.**
All measurements are obtained placing the contacted heterostructure on a piezoelectric stage (*Attocube*), inside a closed-loop Helium cryostat (*Oxford*). Photoluminescence measurements are performed using a homemade confocal microscope (*Olympus* x40 long working distance objective). Spectrally resolved measurements are performed using a supercontinuum laser (*NKT Photonics*) for excitation and a grating spectrometer coupled to a CCD camera (*Andor*) for detection. Time resolved measurements are performed using a pulsed laser diode with tunable repetition rate (*Picoquant*) for excitation and an avalanche photodiode coupled to a time-correlated counter (*PicoHarp*) for detection.

## ACKNOWLEDGMENTS:

F.V. acknowledges financial support from Marie-Curie International Fellowship COFUND and ICFOnest programme. P.S. acknowledges financial support by a scholarship from the 'la Caixa' Banking Foundation. M.M. thanks the Natural Sciences and Engineering Research Council of Canada (PGSD3-426325-2012). T.G.P. is supported by the Danish National Research Foundation, project DNRF103 (CNG) and by the Villum foundation (QUSCOPE). F.H.L.K. acknowledges financial support from the Government of Catalonia through the SGR grant (2017-SGR-1656), and from the Spanish Ministry of Economy and Competitiveness, through the "Severo Ochoa" Programme for Centres of Excellence in R&D (SEV-2015-0522), support by Fundacio Cellex Barcelona and CERCA Programme / Generalitat de Catalunya. Furthermore, the research leading to these results has received funding from the European Union Seventh Framework Programme under grant agreement no. 786285 Graphene Flagship and ERC Synergy Grant Hetero2D. R.J.H. is fully funded by the Graphene NOWNANO CDT (EPSRC Grant No. EP/L01548X/1). M.S. is funded by the EPSRC Grant No. EP/P010180/1. Computer time was provided by Lancaster University's High-End Computing facility. V.I.F. acknowledges support from EPSRC grants EP/S019367/1, EP/P026850/1 and EP/N010345/1.

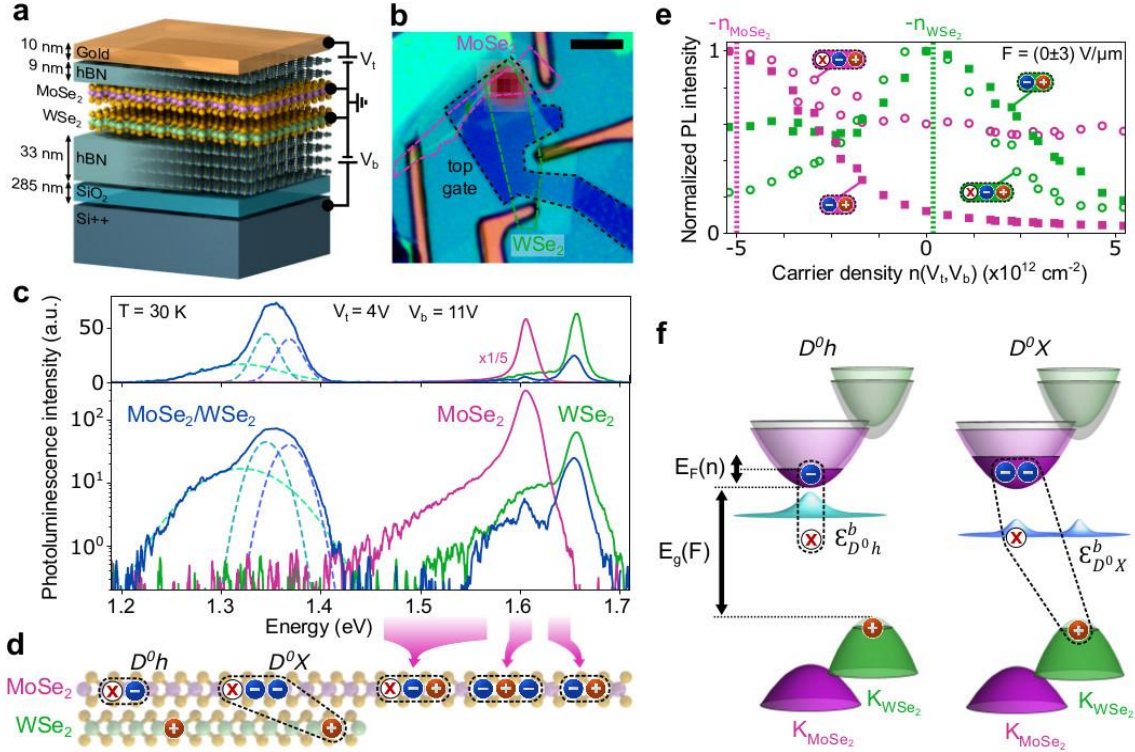

**Figure 1. Impurity-bound interlayer complexes. a.** Schematics of the double-gated hBN/MoSe$_2$/WSe$_2$/hBN van der Waals heterostructure. The backgate consists of doped Si at potential $V_b$, and the topgate is made of a thin Au layer at potential $V_t$. **b.** Real colour optical image of the corresponding fabricated device, where monolayers of MoSe$_2$ and WSe$_2$ and thin top gate are highlighted with purple, green and black dashed lines, respectively. A map of the photoluminescence intensity spectrally integrated between 1.25 and 1.4 eV is superimposed in shades of red. Scale bar is 5 μm. **c.** Photoluminescence spectra (solid lines) on linear (top) and logarithm (bottom) scales, measured on the MoSe$_2$ monolayer (purple), WSe$_2$ monolayer (green) and MoSe$_2$/WSe$_2$ heterobilayer (blue). Dashed lines present the three Gaussian components obtained by fitting the interlayer peak. Light excitation is at energy 2.1 eV, repetition rate 40 MHz and power 3 μW. Temperature is T = 30 K. **d.** Real space schematics of the three intralayer (right) and two interlayer (left) complexes generated in the system. Electrons, holes and impurities are presented as blue, red and white disks, respectively. Dashed lines illustrate Coulomb binding. Purple arrows indicate the spectral ranges corresponding to the intralayer complexes in the MoSe$_2$ luminescence spectrum in c. **e.** Normalized luminescence intensity of MoSe$_2$ (purple) and WSe$_2$ (green) unbound (full squares) and impurity bound (open circles) intralayer exciton peaks as functions of the electrostatically tuned carrier density *n*, at zero field *F*. The carrier densities at the maxima indicate the intrinsic doping of each monolayer (dashed vertical lines). **f.** Bandstructure schematics of the two main interlayer complexes generated in the MoSe$_2$/WSe$_2$ heterobilayer according to *ab initio* calculations. The representation of the carriers is similar to the one in d. Spatial extension of the wavefunctions are represented in light blue within the gap. Relative population of the complexes is tuned through the Fermi energy $E_F$ using carrier density n, while energy of the complexes is tuned through interlayer bandgap $E_g$ using field *F*.

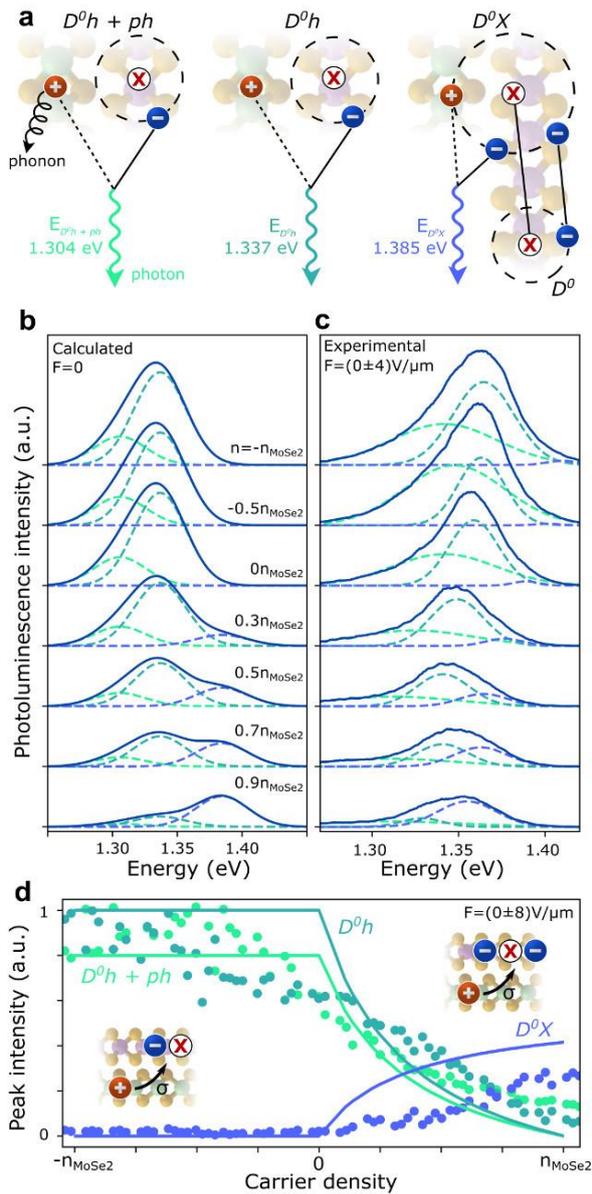

**Figure 2. Carrier density dependence. a.** Diagrams of the three main radiative recombination processes in MoS$_2$/WSe$_2$ described in the text. Representation of carriers is similar to the one in Figures 1d,f. In the background, the WSe$_2$ (green) and MoSe$_2$ (purple) layers illustrate where the carriers reside spatially. **b.** Calculated PL spectra for a range of carrier densities at zero field. A constant empirical 20 meV broadening has been applied to the Gaussian curves. Colours of the three Gaussian components (dashed lines) forming the spectra (solid line) are chosen accordingly with the photons represented in a. **c.** Experimental PL spectra for the same carrier densities as in b, with fixed zero field. The same colour scheme as in b is applied. Experimental conditions are similar to the ones described in Figure 1c. **d.** Experimental (dots) and calculated (line) relative peak intensities, as functions of carrier density. The same colour scheme as in a is applied. Insets: Schematics illustrating the recombination mechanisms of unbound holes with electrons bound to impurities, whose nature varies with carrier density, with a rate that is governed by a cross section $\sigma$ that includes both radiative and non-radiative processes. The representation of carriers and layers is the same as in a.

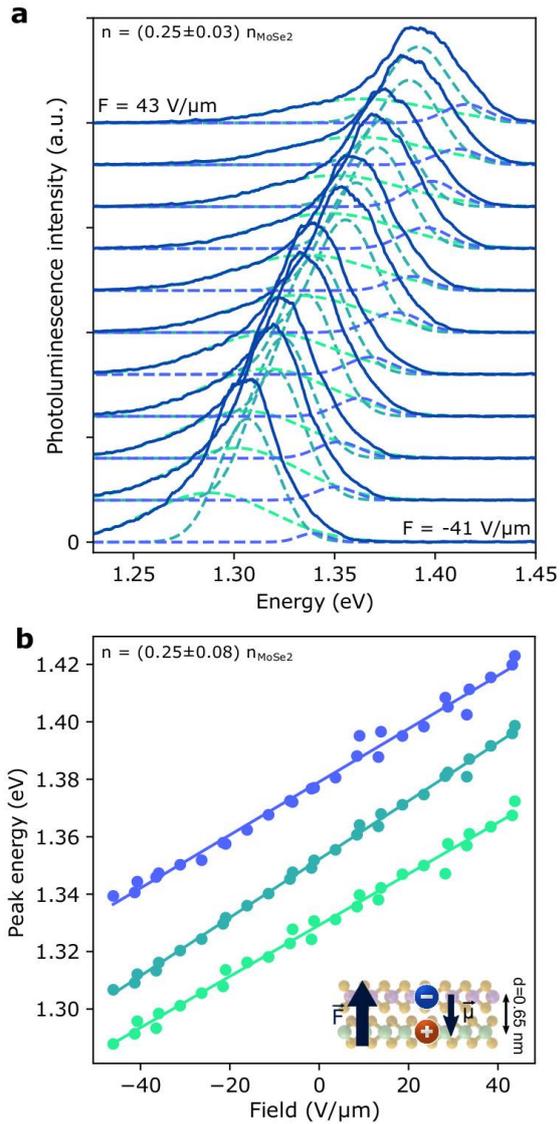

**Figure 3. Out-of-plane field dependence. a.** Experimental PL spectra for a range of out-of-plane fields, with fixed carrier density. The same colour scheme as in Figure 2b is applied. Experimental conditions are similar to the ones described in Figure 1c. **b.** Extracted line energies of the three Gaussian peaks as functions of out-of-plane field, following the same colour scheme. Inset: Schematics of electric field $\vec{F}$ and permanent dipole $\vec{\mu}$ in the system. The representation of carriers and layers is the same as in Figure 1 and 2.

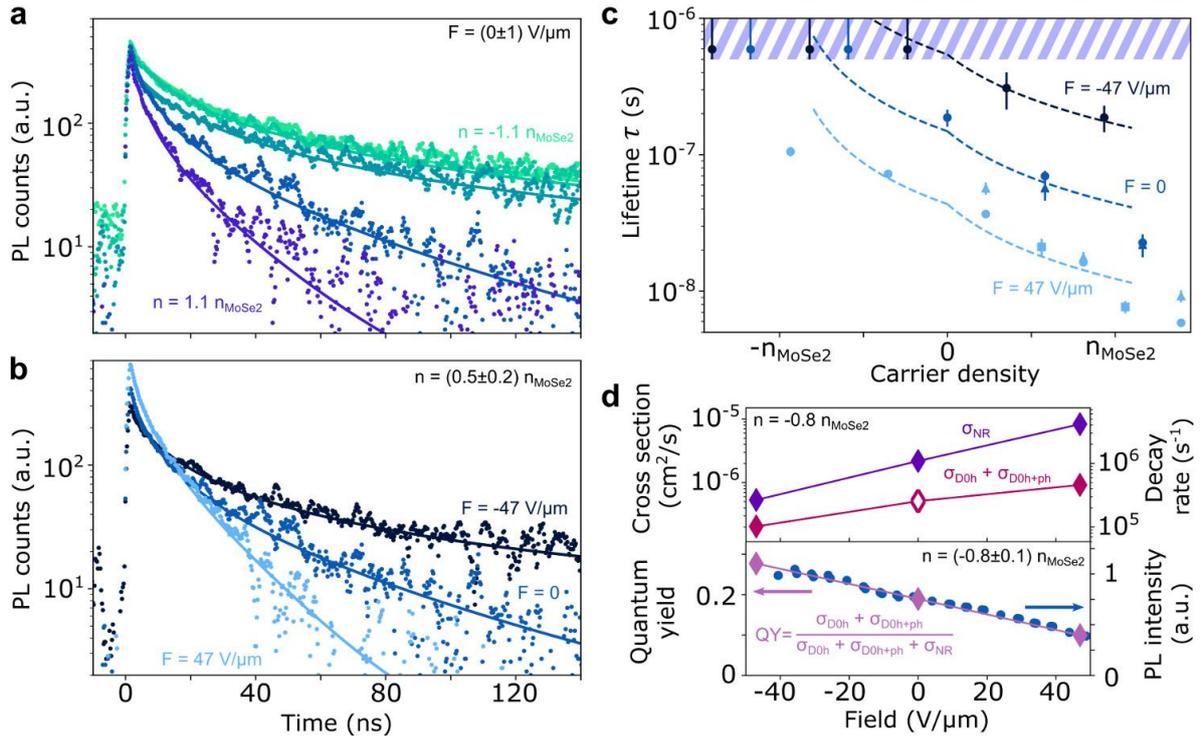

**Figure 4: Time-resolved measurements. a.** Luminescence decay over time at the MoSe$_2$/WSe$_2$ junction for a range of carrier densities, with fixed field. **b.** Similar measurements for a range of fields, with fixed carrier density. In a and b, continuous lines are fits to the experimental data presented as dots using the model discussed in the text and SI. Light excitation is at energy 2.3 eV, repetition rate 2.5 MHz and power 3 μW. **c.** Fitted lifetime as a function of carrier density for a range of fields. Values are extracted from three sets of data obtained with light excitation with different powers and repetition rates, corresponding to initial electron and hole densities of 3 (triangles), 10 (boxes) or 30 (dots) x10$^{12}$ cm$^{-2}$. Unresolved values above 0.5 μs (dashed region) are extracted from the fit at negative carrier densities and negative fields. Dashed lines are fits to the data using the rate model discussed in the text and illustrated in insets of Figure 2d. **d.** (top panel) Cross sections and corresponding decay rates (inverse of $\tau$) at a fixed carrier density, as a function of electric field. (bottom panel) Normalized experimental PL intensity and corresponding fitted emission quantum yield (QY) as function of field for the same fixed carrier density. The non-radiative cross sections (purple diamonds) are extracted from the fitting of the rate model to the data in c. The relative field dependence of the radiative cross sections is obtained from subsequent fitting of the theoretical quantum yield (pink diamonds and inset formula in bottom panel) to the PL data (blue dots). The extracted radiative cross sections (magenta diamonds in top panel) are normalized using the calculated value of the cross section at zero field (open diamond).